\documentclass[baaa]{baaa}

\usepackage[pdftex]{hyperref}
\usepackage{subfigure}
\usepackage{natbib}
\usepackage{helvet,soul}
\usepackage[font=small]{caption}

\contriblanguage{0}

\contribtype{1}

\thematicarea{11}

\title{La inserci\'on de la Astronom\'ia Cultural en la educaci\'on formal: fundamentos y prop\'ositos}

\titlerunning{La inserción de la Astonomía Cultural en la educación}

\author{
J. I. Bastero\inst{1,2}, F. A. Karaseur\inst{1,2}, S. J. Gar\'ofalo\inst{2}
\&
A. Gangui\inst{2,3}
}

\authorrunning{Bastero, Karaseur, Garófalo \& Gangui}

\contact{juanibastero@hotmail.com}

\institute{
Universidad Nacional del Centro de la Provincia de Buenos Aires, Argentina \and
Instituto de Formaci\'on e Investigaci\'on en Ense\~nanza de las Ciencias, Facultad de Ciencias Exactas y Naturales, UBA, Argentina \and
Instituto de Astronom\'ia y F\'isica del Espacio, CONICET--UBA, Argentina
}

\resumen{Son vastos los trabajos de investigación educativa que destacan las serias dificultades que presentan los estudiantes en el aprendizaje de temas astronómicos, así como también en la prevalencia de una enseñanza tradicional distanciada de lo observacional y vivencial, acentuando así las dificultades detectadas. Sostenemos que una enseñanza progresiva con enfoque topocéntrico y contextualizado favorecería la motivación del alumnado, la construcción de una mirada más real de la ciencia actual y un rol más activo en el proceso de aprendizaje. La Astronomía Cultural (AC) es una disciplina académica que busca entender las múltiples formas en las que las sociedades se relacionan con los objetos y fenómenos celestes. Por tal motivo, consideramos que sería un recurso potente para la enseñanza, puesto que brinda herramientas para la contextualización y permite trabajar con experiencias del cielo ligadas a la “astronomía a ojo desnudo”, que requiere de poco o ningún instrumental. Cabe destacar que la AC involucra aspectos de arqueoastronomía, etnoastronomía e historia de la astronomía ofreciendo así múltiples aristas a tener en cuenta. El presente trabajo busca fundamentar la incorporación de estudios en AC para la enseñanza de la astronomía en la educación secundaria y terciaria.}

\abstract{There are vast educational research works that highlight the serious difficulties that students present in learning astronomical subjects, as well as the prevalence of a traditional education distanced from the observational and experiential, thus accentuating the difficulties detected. We argue that progressive teaching with a topocentric and contextualized approach would favor the motivation of the students, the construction of a more real view of current science and a more active role in the learning process. Cultural Astronomy (CA) is an academic discipline that seeks to understand the multiple ways in which societies relate to celestial objects and phenomena. For this reason, we consider that it would be a powerful resource for teaching, since it provides tools for contextualization and allows working with sky experiences linked to “naked eye astronomy", which requires little or no instruments. It should be noted that CA involves aspects of archaeoastronomy, ethnoastronomy and the history of astronomy, thus offering multiple dimensions to take into account. The present work seeks to base the incorporation of CA studies for astronomy teaching in secondary and tertiary education.}


\begin{document}

\maketitle

\section{Introducción}\label{S_intro}

Diversos trabajos de investigación en didáctica de las ciencias señalan varias dificultades al momento de aprender temas astronómicos en la escuela y en la formación de docentes \citep{Camino_1995, Gangui_2010, DeBiasi_2015, Varela_2015}. A su vez, estos trabajos también señalan cómo, frecuentemente, tanto alumnos como docentes presentan concepciones alternativas de los fenómenos astronómicos cotidianos como lo pueden ser las estaciones del año, el paso cenital del Sol o las fases lunares, entre otros.
En astronomía los alumnos suelen contar con mucha información sobre las estrellas y sus movimientos \citep{Lopez_Scarinci_2006}, pero no logran conectar dicha teoría con lo que experimentan en su vida cotidiana con el fin de crear un modelo mental coherente para explicar los fenómenos diarios.
Parte de estas dificultades se hallan estrechamente relacionadas a la forma en la que se imparten los conocimientos en las aulas. En la enseñanza de estos temas generalmente persiste la transmisión de contenido desde un observador externo y sin relacionarlo con el mundo vivencial, lo que conduce a construir conocimientos fragmentados \citep{Faria_Volke_2008}. Este enfoque de enseñanza, que dista de la experiencia cotidiana del alumnado, contribuye a acentuar varias de las concepciones alternativas. En contraparte, una mirada topocéntrica de estos fenómenos permitiría al docente contextualizar los conocimientos generando así modelos de ciencia escolar \citep{AdurizBravo_IzquierdoAymerich_2009, Gangui_Iglesias_2015}.

\section{La Astronom\'ia Cultural}

La Astronomía Cultural, término acuñado por Iwaniszewski y Ruggles en la década del 1990, intenta establecer las concepciones sobre el cielo que han ido forjando las personas de diversas culturas, las preguntas que se hicieron, las respuestas que han dado y cómo han evolucionado con el tiempo las mismas. En ella, se busca entender las distintas formas en las que los objetos y fenómenos del cielo se registran, influyen, impactan y guían las creencias, los sistemas de conocimiento y las tradiciones culturales.
La Astronomía Cultural es un área interdisciplinaria, desde la base de las ciencias naturales y las sociales, que presenta diversas aplicaciones en áreas como la educación, las artes, la política, etc. Dentro del gran espacio que conforma, tres de las subdisciplinas principales son \citep{Lopez_Hamacher_2017}:

$\bullet$  La arqueoastronomía, que, mediante el uso de las técnicas de la arqueología y de la astronomía, reconstruye las formas con que, en el pasado, distintos grupos humanos vieron el cielo. 

$\bullet$  La etnoastronomía, que mediante una aproximación etnográfica, intenta entender las concepciones sobre lo celeste que tienen los diversos grupos étnicos y culturales.

$\bullet$  La historia de la astronomía, que se dedica al estudio de la astronomía del pasado mediante el uso de las técnicas históricas y el soporte de los documentos escritos.

\section{La Astronom\'ia Cultural como recurso did\'actico}

Distintas investigaciones en enseñanza de las Ciencias Naturales respaldan que en las clases se enfaticen las relaciones entre contenidos científicos, aspectos socioculturales, de naturaleza de la ciencia y de la vida cotidiana de los estudiantes con el fin de lograr aprendizajes significativos y competenciales \citep{Caamanno_2018, Habig_2018}. Estos aspectos convergen cuando se aborda la enseñanza mediante un enfoque que permita la modelización y contextualización al generar un entorno motivador que mejora los procesos de comprensión \citep{Pergola_Galagovsky_2020}. 
En esta línea teórica, la ciencia puede ser pensada como una actividad humana y el conocimiento construido, como una familia de modelos científicos que dan cuenta de distintos fenómenos del mundo \citep{Meinardi_2002}. La importancia de incorporar a la enseñanza la Astronomía Cultural radica en que abre un espacio a la reflexión acerca de las formas en que ese conocimiento se construye socio-culturalmente. Este marco permite retomar las ingeniosas preguntas que se hacían y aún se hacen distintas culturas como puertas de entrada estratégicas para la enseñanza. 
Este enfoque histórico-epistemológico acerca a los estudiantes a conocer no sólo cómo la humanidad fue construyendo los modelos científicos actuales, sino también cómo dicho conocimiento es siempre un conocimiento “situado”, ligado a quienes somos \citep{Lopez_Hamacher_2017}. 
Teniendo en mente el potencial didáctico que tiene la incorporación de la Astronomía Cultural para la educación formal se diseñaron dos propuestas de enseñanza. Si bien ambas fueron pensadas para ser implementadas en Ciencias Naturales de secundaria, las dos pueden ser adaptadas también para ser llevadas al aula en institutos de formación docente. 

En una de ellas, se aborda la enseñanza de conceptos de espacio y tiempo utilizando el caso de la navegación oceánica histórica sin instrumentos avanzados, apelando a estudios etnoastronómicos. En la otra propuesta, mediante una aproximación arqueoastronómica, con el ejemplo del calendario de horizonte monumental del sitio arqueológico Chanquillo, se aborda la identificación local de marcadores de horizonte que permitiría a los estudiantes la construcción de sus propios calendarios. Ambas se describen con mayor profundidad en una segunda parte que da continuidad a este trabajo (Karaseur et al., trabajo aceptado en BAAA 63, 2021).

\section{Reflexiones finales}

En el presente trabajo se describe el potencial de la Astronomía Cultural como recurso didáctico con el propósito de orientar al docente hacia la posibilidad de abordar el cambio de enfoque necesario que permita promover la modelización, la asociación entre los contenidos disciplinares y los aspectos socioculturales, y cuestiones de la vida cotidiana de los estudiantes. Así mismo, esta forma de intervención en el aula estimula, desde la enseñanza, el abordaje hacia nuevos procedimientos observacionales y formas vivenciales de concebir la exploración de los cielos \citep{Lopez_GimenezBenitez_2010}. 


\bibliographystyle{baaa}
\small
\bibliography{739_v2}
 
\end{document}